\documentclass[conference]{IEEEtran}
\IEEEoverridecommandlockouts
\usepackage{cite}
\usepackage{amsmath,amssymb,amsfonts}
\usepackage{url}
\usepackage{graphicx}
\usepackage{textcomp}
\usepackage{xcolor}
\usepackage{algorithm,algorithmic}
\usepackage{float}
\usepackage{tabularx}
\usepackage{stfloats} 
\usepackage{fancyhdr}

\fancypagestyle{customhead}{
    \fancyhead{} 
    \fancyfoot{} 

    \fancyhead[C]{\textit{\footnotesize This work is presented as a conference paper at the 2025 IEEE Frontiers in Education 52nd Conference (FIE)}}
}

\def\BibTeX{{\rm B\kern-.05em{\sc i\kern-.025em b}\kern-.08em
    T\kern-.1667em\lower.7ex\hbox{E}\kern-.125emX}}

\begin{document}

\pagestyle{customhead} 



\title{Bridging the Gap Between Theoretical and Practical Reinforcement Learning in Undergraduate Education}

\author{
\and
\IEEEauthorblockN{Muhammad Ahmed Atif}
\IEEEauthorblockA{\textit{Dhanani School of Science and Engineering} \\
\textit{Habib University}\\
Karachi, Pakistan \\
muhammad.atif@habib.edu.pk}
\and
\IEEEauthorblockN{Mohammad Shahid Shaikh}
\IEEEauthorblockA{\textit{Dhanani School of Science and Engineering} \\
\textit{Habib University}\\
Karachi, Pakistan \\
shahid.shaikh@sse.habib.edu.pk}
}

\maketitle

\thispagestyle{customhead}

\begin{abstract}
This innovative practice category paper presents an innovative framework for teaching Reinforcement Learning (RL) at the undergraduate level. Recognizing the challenges posed by the complex theoretical foundations of the subject and the need for hands-on algorithmic practice, the proposed approach integrates traditional lectures with interactive lab-based learning. Drawing inspiration from effective pedagogical practices in computer science and engineering, the framework engages students through real-time coding exercises using simulated environments such as OpenAI Gymnasium. The effectiveness of this approach is evaluated through student surveys, instructor feedback, and course performance metrics, demonstrating improvements in understanding, debugging, parameter tuning, and model evaluation. Ultimately, the study provides valuable insight into making Reinforcement Learning more accessible and engaging, thereby equipping students with essential problem-solving skills for real-world applications in Artificial Intelligence.

\end{abstract}


\maketitle

\section{Introduction}
Reinforcement learning (RL) is a fundamental paradigm in artificial intelligence (AI), where agents learn to make sequential decisions by interacting with an environment to maximize a notion of cumulative reward over time. Based on concepts from behavioral psychology, RL enables agents to navigate complex and uncertain environments through trial-and-error learning, characterized by an exploration-exploitation trade-off that adjusts behavior based on the outcomes of past actions \cite{sutton_barto}. In RL, agents operate in a Markov Decision Process (MDP) framework, where they perceive states, take actions, and receive rewards, optimizing their strategy, or ``policy," to achieve the highest cumulative reward \cite{mdp-richard}. Core RL methods such as as Q-learning, temporal difference (TD) learning, and policy gradient methods \cite{sutton_barto,watkins1992q} laid the foundation for applications in fields as varied as robotics, healthcare, finance, and gaming. The advent of deep RL, which combines deep neural networks with RL framework, further accelerated the field, as demonstrated in landmark systems such as DeepMind's AlphaGo \cite{bbc2016alphago}, where agents reached superhuman performance levels by autonomously developing sophisticated strategies in real time \cite{mnih2015human}.

The growth and development of RL has made this field more mainstream in education and has prompted significant innovation in its teaching methodologies, reflecting an understanding that theoretical knowledge must be augmented with practical application to fully grasp RL's dynamic learning processes. A typical RL education often emphasized the MDP framework, Bellman equations, and convergence proofs, which, while essential, limited students' ability to intuitively understand agent behavior and policy improvement in real-world scenarios. Today, research increasingly supports the integration of interactive hands-on methods into RL education \cite{kubotani2021rltutor,moerland2023edugym}. Platforms like OpenAI Gymnasium \cite{towers2024gymnasium} and Unity ML-Agents \cite{juliani2018unity} provide students with accessible, standardized environments to experiment with RL algorithms, observe agent decision-making, and understand the impacts of tuning hyper-parameters on performance stability and learning rates \cite{mnih2015human}. These environments facilitate visualization of agent learning and help clarify complex RL mechanism, such as balancing exploration with exploitation, managing the stability-plasticity dilemma \cite{rudroff2024neuroplasticity}, and handling non-stationarity in environments.

General pedagogical advancements in engineering and computer science also reflect a shift toward active learning through simulations and visualization tools \cite{inproceedings}. This shift is driven by a growing recognition that hands-on, interactive experiences enable students to grasp complex theoretical concepts more effectively than traditional lecture-based approaches. These technologies facilitate interactive and immersive learning experiences that allow students to engage deeply with complex concepts. For instance, the use of program visualization in technology-constrained classrooms has been shown to enhance student engagement and comprehension, even when direct interaction with the visualization is mediated by the instructor \cite{banerjee2015effect}. These approaches emphasize a significant shift toward active learning paradigms, leveraging simulations and visualization tools to improve educational outcomes in STEM fields.

Instructors are now leveraging educational frameworks such as project-based learning and collaborative coding to emphasize iterative debugging, a skill essential to RL due to the difficulty in diagnosing performance issues in complex environments. Furthermore, the integration of machine learning frameworks such as TensorFlow \cite{tensorflow2015-whitepaper} and PyTorch \cite{paszke2019pytorchimperativestylehighperformance} in RL courses enables students to design, implement and test RL algorithms from scratch, fostering a deeper understanding of architectural choices and computational constraints involved in developing and deploying RL systems. As RL continues to evolve with applications in autonomous systems, adaptive control, and dynamic systems modeling, educational research establishes the need for an adaptive curriculum that evolves alongside emerging RL techniques.

The pedagogical innovations in RL teaching should address the dual emphasis on foundational knowledge and experiential learning, preparing students to tackle real-world problems with a nuanced understanding of RL’s strengths and limitations. As RL systems become more complex and pervasive, ongoing research in RL pedagogy remains critical to advancing not only technical proficiency but also ethical, interpretive, and critical-thinking skills among future practitioners.

\section{Literature Review} 
While there are not many prior pedagogy-specific RL studies, reviewing literature from broader researches is certainly helpful in understanding the pedagogical gaps in RL education and bridging the gap between theoretical and practical RL education. Across most pedagogical research and studies, the theory and praxis of teaching approaches and strategies constitute a mutually reinforcing framework in which rigorous theoretical models inform the design of instructional practices, and praxis involves the reflective, iterative refinement of these practices in authentic educational settings. Within this paradigm, foundational learning theories, including behaviorism and cognitivism \cite{dilshad2017learning}, initially focused on observable behavior and mental processes, respectively. Alongside constructivist perspectives which assert that learners actively build knowledge through experience and reflection \cite{bransford2005learning}, the Vygotsky’s notion of the Zone of Proximal Development highlights the critical role of social interaction and scaffolding in facilitating learning \cite{proximal} which promote independent learning \cite{9225989}. These theoretical insights underpin many contemporary pedagogical strategies, particularly in undergraduate and laboratory-based teaching environments.

Within undergraduate education, especially in STEM disciplines, laboratory-based instruction is indispensable for bridging the gap between abstract theoretical constructs and practical application. Undergraduate science and engineering curricula often combine traditional lectures with laboratory experiments and project-based activities that replicate real-world challenges, thereby reinforcing theoretical concepts through direct, hands-on experience. Such pedagogical methods are designed to foster active learning, critical thinking, and collaborative problem solving. Praxis in these contexts is realized through a systematic cycle of planning, implementation, evaluation, and adjustment, whereby educators continuously refine laboratory exercises and instructional methods based on student performance and feedback. This dynamic interplay between theory and praxis not only deepens conceptual understanding but also cultivates the essential skills required for professional practice in complex, real-world environments \cite{4720628}.

 Recent studies have shown that incorporating project-based learning (PBL) and agile methodologies in the curriculum fosters a collaborative and iterative learning environment \cite{PBL}. In such settings, students work on real-world problems, developing, testing, and refining their solutions in a manner that closely mirrors industry practices. This approach not only deepens their understanding of complex algorithms and systems but also cultivates essential skills in teamwork, adaptive thinking, and self-regulated learning. Empirical evidence from research on using PBL and agile techniques to teach AI confirms that these methods align with the cognitive apprenticeship model \cite{collins2006cognitive}, thereby facilitating the transition from novice to expert performance in computing education.

Furthermore, the emerging discipline of reinforcement learning (RL) stands to benefit significantly from advances in Computer Science (CS) and AI pedagogy. Like many traditional CS subjects, RL involves complex problem-solving and demands that learners integrate theoretical knowledge with practical, iterative application. In other words, successful RL instruction must move beyond mere lecture-based delivery of mathematical and algorithmic content and should include active, hands-on learning experiences that build intuition over time.

Textbooks and lectures in RL often present abstract equations and algorithmic concepts, and public codebases and research prototypes may be too complex for beginners. However, recent tools such as EduGym \cite{moerland2023edugym} and RLTutor \cite{kubotani2021rltutor} have emerged as promising solutions. EduGym offers a suite of RL environments paired with interactive notebooks that directly connect theoretical equations to code examples. In parallel, RLTutor uses a virtual student model to optimize teaching strategies with minimal interaction, serving as a buffer between theoretical instructional optimization and practical e-learning applications. RL-Lab \cite{Salloum2021RL-Lab} is another promising platform in making RL education more accessible and intuitive. The idea of RL-Lab is to allow students to experience and play with RL before they write a single line of code. It enables people to design their own situations, select and parameter algorithms, execute the learning process, and determine what went well and what didn't based on the results.

Zendler demonstrates that aligning teaching methods with underlying learning theories—behavioristic, cognitivist, and constructivist—can significantly enhance course design and student engagement \cite{zendler2019teaching}. Similarly, Hazzan, Ragonis, and Lapidot emphasize the value of active-learning strategies such as CS-Unplugged activities, collaborative problem solving, and project-based learning \cite{hazzan2020teaching}. These approaches encourage students to actively apply theoretical principles in controlled, yet realistic, settings. In RL courses, where students must work with complex concepts such as the Bellman equations, exploration–exploitation trade-offs, and policy evaluation, such active-learning practices help students develop a deeper, more intuitive understanding that supports the practical coding and experimentation necessary for real-world success.

\section{Overview of Reinforcement Learning}
Reinforcement Learning enables an agent to learn optimal behaviors through a process of trial and error, guided by feedback from its environment. A classic analogy is a child learning to ride a bicycle: the child (agent) interacts with the cycling road (environment) and receives rewards, such as, staying upright (positive reward) or falling (negative reward). This interaction exemplifies the core RL mechanism, where actions are evaluated based on their outcomes to refine future decisions. The RL framework formalizes this process, into an agent, an environment, and a reward system. The goal is to optimize a policy function that dictates the agent’s actions given current state to maximize cumulative rewards. A general RL framework is illustrated below:

\begin{figure}[H]
  \centering
  \includegraphics[width=0.45\textwidth]{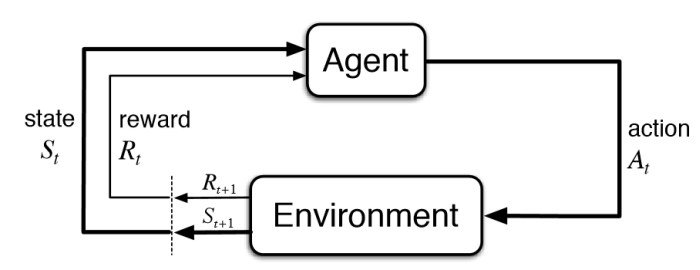}
  \caption{From Sutton and Barto \cite{sutton_barto}}
  \label{fig:6}
\end{figure}

Mathematically, RL relies on the Markov Decision Processes (MDPs) which extends the Markov Chain to models by incorporating decision making, provides the mathematical framework for RL \cite{sutton_barto}. An MDP is characterized by states, actions, transition probabilities, and rewards, forming the foundation for defining Value functions. These Value functions, which predict the expected cumulative reward from a given state or state-action pair, are computed using the Bellman equation which serves as a recursive formulation to determine the optimal policy by evaluating the trade-offs between immediate and future rewards. The Bellman equations for the state-value function, $V^*$, and the action-value function, $Q^*$, in an MDP, respective, are:
$$
V^*(s) = \max_{a} \left\{  \sum_{s'} \left(\sum_{r} P(s', r \mid s, a) \left[ r + \gamma V^*(s') \right] \right) \right\}
$$
and
$$
Q^*(s, a) = \sum_{s'} \left[ \sum_{r} P(s', r \mid s, a) \left( r + \gamma \max_{a'} Q^*(s', a') \right) \right]
$$
Where:
\begin{itemize}
\item  $s$ is the current state,
\item  $a$ is the action,
\item  $s'$ is the next state,
\item  $r$ is the immediate reward, and
\item  $P(s', r \mid s, a)$ is the probability of transitioning to the state $s'$ and receiving a reward $r$ where action $a$ is taken in the current state $s$, and
\item  $\gamma$ is the discount factor $(0 < \gamma \leq 1)$.
\end{itemize}

Different methods are used to solve these Bellman equations, each offering their own advantages and disadvantages. Some of the popular and widely taught techniques are Dynamic Programming \cite{bertsekas2019reinforcement}, Monte-Carlo methods \cite{hammersley2013monte} and Temporal Difference methods \cite{tesauro1995temporal}.

RL techniques are categorized into model-free and model-based methods. Model-free methods, such as Q-learning and Deep Q-Networks (DQN), directly learn value functions and policies, from interactions with the environment, without constructing a model of the transition dynamics. These approaches are computationally simpler but may require extensive exploration. Conversely, model-based RL builds an internal model to predict state transitions and rewards, enabling efficient planning and policy optimization. Model-based techniques such as Model Predictive Control (MPC) \cite{gorges2017relations} and Dyna-Q \cite{zou2020pseudo} integrate learning and planning to balance exploration and exploitation.

Prediction and Control are central to RL, with prediction methods focusing on estimating value functions to evaluate current policies, while control methods optimize the policy itself. Temporal Difference (TD) learning, which combines the strengths of Monte Carlo methods and dynamic programming, is widely used for this purpose. 

Deep Reinforcement Learning (Deep RL) enhances RL’s scalability to high-dimensional problems by leveraging neural networks to approximate value functions and policies. DQN, for instance, employed convolutional networks to achieve human-level performance in complex environments like Atari games\cite{mnih2015human}. 

Advanced policy gradient methods, including Proximal Policy Optimization (PPO) \cite{schulman2017proximal} and Trust Region Policy Optimization (TRPO) \cite{schulman2015trust}, further improve stability and efficiency in policy optimization.

In more advanced RL, there are more sophisticated methods such as Actor-critic Methods \cite{grondman2012survey}. Actor-Critic methods integrate value-based and policy-based approaches, represent a sophisticated RL architecture. The actor updates the policy by selecting actions, while the critic evaluates these actions by estimating value functions, thus trading off variance reduction of policy gradients with bias introduction from value function methods \cite{williams1992simple} and optimize the policy by adjusting it in the direction of higher expected rewards. Actor-critic variants such as Advantage Actor-Critic (A2C) and Asynchronous Advantage Actor-Critic (A3C) reduce variance in policy gradient estimates, enhancing convergence and learning efficiency in dynamic environments.

\section{Course Syllabus/Design}
The syllabus presented here was initially designed as a 300 level Spring semester 2023 course at Habib University. The course was designed for undergraduate computer science and engineering students and had only the lecture component. Since the first iteration some modifications have been made in response to student course evaluation and some more advanced topics had been added to the syllabus as well. The Course Learning Outcomes are defined as following:
\begin{enumerate}
    \item  Define reinforcement learning and distinguish it from other machine learning paradigms.
    \item  Develop models of some well-known machine learning problems using the RL framework.
    \item  Describe and analyze some well-known algorithms for solving problems given in the standard RL framework.
    \item Implement common RL algorithms using Python programming language and off-the-shelf solutions.
\end{enumerate}
    
\subsection{Lecture Component}
The Lecture Component primarily covers RL concepts with the focus on mathematical rigor and depth. A considerable time is spent in understanding the mathematical formulation of RL and how problems in RL can be translated to Markov-Decision Processes. This helps students in the understanding of more advanced topics. The lecture component also focuses on how popular algorithms, such as Q-Learning, are derived. A Wide range of topics, listed below, are covered in the lecture from the textbook \cite{sutton_barto}:
\begin{itemize}
    \item  Elements of Reinforcement Learning (Chapter 1),
    \item The standard Reinforcement Learning framework (Chapter 1),
    \item  Markov decision processes (Chapter 3),
    \item  Policy evaluation, improvement and Policy Iteration (Chapter 4),
    \item  Dynamic Programming and Value iteration (Chapter 4),
    \item Monte-Carlo Methods (Chapter 5),
    \item Temporal-Difference (TD) Method (Chapter 6),
    \item Q-Learning and SARSA (Chapter 6),
    \item Approximate methods and Deep Q-Networks (Chapter 9),
    \item Policy Gradient and Actor-Critic methods (Chapter 13).
\end{itemize}
\subsection{Lab Component}
The lab component, on the other hand, primarily focuses on the high-level implementation of the concepts. Each lab is designed as a Jupyter notebooks, featuring elaborate descriptive text coupled with coding exercises. 

The notebooks are designed in such a manner that a student can understand a particular concept at a high-level. Each notebook focuses on developing a thorough understanding of where these concepts are relevant and how to breakdown a certain class of problems into the framework of RL. Moreover, the notebook guides and develops relevant skills needed to implement RL based solutions for those problems by hands-on coding exercises and challenges. The labs are implemented following the ``Scaffolding with fading" strategy \cite{collins2006cognitive}, which is a key component of the cognitive apprenticeship model. In this approach, the instructor initially provides significant support (scaffolding) and gradually withdraws that support (fading) as learners gain competence. This method helps students become increasingly independent and gain confidence in their understanding of the subject through structured assessments and practice.

Certain labs are designed to allow students to prepare a trained agent for a particular environment that will compete against the trained agent of other students, in a tournament like fashion. The motivation of developing those labs in such manner is to make the course and its contents more engaging, student-centric and fun.

The programming language of choice is python. Various libraries are explored alongside the python implementations such as OpenAI Gymnasium \cite{towers2024gymnasium}, TensorFlow \cite{tensorflow2015-whitepaper} and so on.

Following is the list of topics covered in each lab notebook:
\begin{itemize}
    \item \textbf{Lab 01:} Introduction to RL and OpenAI Gymnasium,
    \item \textbf{Lab 02:} MDP and its solution using Dynamic Programming,
    \item \textbf{Lab 03:} Monte-Carlo Methods,
    \item \textbf{Lab 04:} Temporal Difference Learning,
    \item \textbf{Lab 05:} Neural Networks Recap,
    \item \textbf{Lab 06:} TensorFlow Recap,
    \item \textbf{Lab 07:} Deep Q-Network (DQN),
    \item \textbf{Lab 08:} Double and Dueling DQN,
    \item \textbf{Lab 09:} Actor-Critic Methods, and
    \item \textbf{Lab 10:} Advanced RL.
\end{itemize}

\section{Teaching Approaches}
Reinforcement Learning Education is in a position where its highly theoretical lectures need a counter-balance of intuition building labs where Reinforcement Learning's relevance and practicality is realized in a more intuitive manner. Integrating theory and praxis is paramount due to the inherently complex and abstract nature of RL algorithms. Reinforcement learning is deeply rooted in mathematical formulations—such as Markov Decision Processes, value functions, and temporal-difference learning—that require a robust theoretical foundation as well as extensive practical experimentation. Seminal works by Sutton and Barto \cite{sutton_barto} provide a comprehensive theoretical framework that is essential for understanding RL concepts, while studies demonstrate how these principles are applied in cutting-edge research \cite{article}. In undergraduate curricula, laboratory-based exercises, such as simulations using platforms like OpenAI Gymnasium or custom-designed projects, offer students the opportunity to implement RL algorithms in controlled, experimental settings. Such hands-on activities facilitate the experiential learning process, allowing students to iteratively test, refine, and internalize abstract concepts.

This pedagogical approach aligns with cognitive apprenticeship models, as articulated by Collins, Brown, and Newman \cite{inbook}, which emphasize the importance of scaffolding, modeling, and the gradual fading of support. By initially providing structured guidance in laboratory settings such as RL-Labs \cite{Salloum2021RL-Lab} and then progressively encouraging independent problem solving using more advanced framework like Tensorflow, Pytorch and Gymnasium, educators can effectively bridge the gap between theoretical instruction and practical application. Moreover, Vygotsky’s concept of the Zone of Proximal Development \cite{mindinsociety} highlights the importance of collaborative and scaffolded learning environments in fostering deep conceptual understanding. The combination of these theoretical insights with rigorous lab-based practices not only enhances students’ mastery of reinforcement learning techniques but also cultivates critical problem-solving skills that are vital for both academic research and industry applications.

Another teaching strategy employed in Spring semester 2025 RL course iteration was the use of mix-assignment structure in which half of the assignment is coding based and another half is theory based. The way these two sections are structured is that the theory part is dependent and builds on the coding section. In this way students are made to approach and think through the assignment problems by establishing concrete relationship and relevant between practical and theoretical solution and approaches.

\section{Survey}
\begin{table*}[t]
\centering
\caption{Survey for Students of Spring 2025}
\begin{tabular}{|c|l|c|}
\hline
\textbf{No.} & \textbf{Question} & \textbf{Type} \\ 
\hline
1 & Have you heard of Reinforcement Learning (RL) before? & Boolean \\
\hline
2 & Where did you get to know about RL? & Multiple Choice \\
\hline
3 & If you answered ``Yes,'' how familiar are you with RL? & Scale/Rating \\ 
\hline
4 & If you are familiar with RL, how confident are you that you understand the basic concepts of RL? & Scale/Rating \\
\hline
5 & How interested and motivated are you to learn more about RL? & Scale/Rating \\
\hline
6 & If you are familiar with RL, how confident are you that you can solve small RL problems? & Scale/Rating \\
\hline
7 & Would you prefer having a lab component in the RL course? & Scale/Rating \\ 
\hline
8 & Would you be more interested in learning the Theory of RL or the implementation of RL? (Theory, Implementation, Both) & Multiple Choice \\
\hline
9 & In RL, what kind of Project would you prefer?  (Research based, Competition based, Hardware based) & Multiple Choice \\
\hline
10 & As a student, what teaching style do you prefer? (Traditional lecture style, Hands-on-learning approaches, Flipped Class) & Multiple Choice \\
\hline
11 & As a student, do you prefer to do projects in a group? & Boolean \\
\hline
12 & As a student, which assessment style do you prefer? (Peer-teaching, Class Activities, Traditional Quizzes) & Multiple Choice \\
\hline
\end{tabular}
\label{table: current cohort}
\end{table*}

\subsection{Data Collection and Assessment} 
This study employed a mixed-methods approach to evaluate the impact of integrating interactive labs into the reinforcement learning (RL) curriculum. Data were collected through complementary primary and secondary sources to enable robust longitudinal analysis. Primary data were gathered via:
\begin{enumerate}
    \item Google Forms survey administered to students enrolled in the Spring 2025.
    \item Students' gradebook of Spring 2025 course iteration.
    \item Students' gradebook of Spring 2023 course iteration (control).
\end{enumerate}
Google Forms survey administered to students enrolled in the beginning of Spring 2025 offering of Introduction to Reinforcement Learning ($N = 26$). The survey combined Likert-scale \cite{likert1932} items and open-ended questions to quantitatively assess perceptions of pedagogical interventions (e.g., interactive labs) and qualitatively capture detailed feedback on students’ experiences. It further evaluated preferences for theoretical versus practical learning, perceived effectiveness of labs, and familiarity with RL concepts, while probing prior exposure to RL through academic, social media, or informal channels. Anonymized data collection protocols ensured participant confidentiality.

The students' gradebook were used to evaluate their performance in the course and gauge the effectiveness of pedagogical interventions used in Spring 2025, compared to Spring 2023 which is used as a control group.

Secondary data were sourced from 2 repositories: 
\begin{enumerate}
    \item An internal course evaluation form completed by the Spring 2025 cohort. 
    \item An internal course evaluation form completed by the Spring 2023 cohort (control).
\end{enumerate}

The Spring 2023 dataset ($N = 28$), which comprises of both quantitative students grades and qualitative internal course evaluation form, served as a control group. It establishes a historical baseline for comparative analysis of learning outcomes, critical thinking development, and curricular effectiveness. The internal course evaluation provided rich quantitative and qualitative insights, with students rating dimensions such as syllabus clarity, course materials, and instructor effectiveness on a 5-point Likert scale \cite{likert1932}. Results from 2023 indicated consistently high ratings (often exceeding 4/5), particularly for the syllabus structure, Canvas course organization, and constructive assignment feedback. Qualitative comments highlighted strengths like the rigorous mathematical foundation and engaging assessments, while suggesting areas for improvement, such as expanded reference materials and enhanced practical applications to bridge theory and practice.

Four interconnected instruments facilitated a comprehensive assessment:

\begin{itemize}
    \item \textbf{Spring 2025 Course Evaluation:} Established baseline metrics for curriculum delivery and student satisfaction.
    \item \textbf{Spring 2023 Course Evaluation:} Established baseline metrics for curriculum delivery and student satisfaction.
    \item \textbf{Spring 2025 RL Survey:} Focused on current perceptions of pedagogical innovations, prior RL exposure, and conceptual familiarity.
    \item \textbf{Student Performance Data:} Grades from assessments (e.g., exams, quizzes) in both semesters enabled quantitative comparison of learning outcomes.
\end{itemize}

By triangulating these datasets, the study enabled a rigorous evaluation of the pedagogical interventions. The longitudinal design allowed direct comparison between the 2023 baseline course iteration and 2025 course iteration, assessing shifts in critical thinking, practical skills, and engagement. Convergence of quantitative metrics (e.g., grade trends, Likert scores \cite{likert1932}) and qualitative feedback (e.g., open-ended responses) provided nuanced insights into the efficacy of interactive labs. This dual-pronged strategy—combining validated surveys, archival evaluations, and performance data—ensured a holistic understanding of how curriculum innovations influenced RL education, while informing future adjustments to align with evolving demands in AI pedagogy.

\begin{table*}[h!]
\centering
\caption{Comparative analysis of students' performance in both semesters \\ (expressed as percentages)}
\begin{tabular}{|c|c|c|}
\hline
\textbf{Statistic} & \textbf{2025 Semester} & \textbf{2023 Semester} \\ 
\hline
Mean &	88.31 &	80.7\\
\hline
Median & 90.09 & 83.10 \\
\hline
Standard Deviation & 7.92	& 16.47 \\ 
\hline
Skewness & 	-2.57 (Left Skew)	& -3.02 (Left Skew) \\
\hline
CI (95\%) & 85.40 – 91.21& 74.38 – 87.16 \\
\hline
\end{tabular}
\label{table: Descriptive Stats}
\end{table*}

\subsection{Survey Result}
The course evaluation feedback from the Spring 2023 semester  was generally positive, with the course receiving an overall score of 4.48 out of 5. However, there was considerable qualitative feedback that strongly suggested integrating more practical elements in the course. Some of the qualitative feedback were as follows.
\begin{itemize}
    \item \textit{``More practical work. better planning."
    \item  ``The content can be made slightly more challenging to make assignments and projects more enjoyable."
    \item ``Including real life examples of RL agents or applying learnt RL methodologies to real life scenarios."
    \item ``The course has very interesting content. however, there are not a lot of reference materials which can assist students in understanding the concepts."
    \item ``Needs to be more practical and not just theoretical. quizzes need to be something very close to the textbook content."}
\end{itemize}

With respect to Spring 2025 RL Survey, the response to Question $8$ (Fig \ref{fig: tableq8}) of Table \ref{table: current cohort}, implied that students generally want a lab component in a RL course with $65.4\%$ students voting to learn both theoretical and practical elements and $3.6\%$ more students preferring pure practical component over pure theoretical component in this course:
\begin{figure}[H]
  \centering
  \includegraphics[width=0.45\textwidth]    {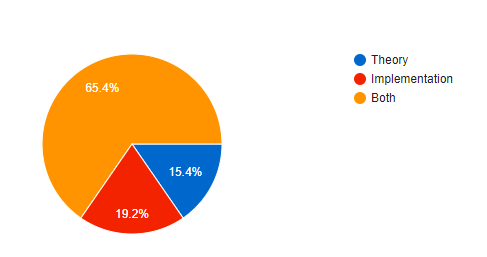}
  \caption{Response to the question: ``Would you be more interested in learning the Theory of RL or the Implementation of RL?" from Table \ref{table: current cohort}}
  \label{fig: tableq8}
\end{figure}

Another strong indicator of integrating lab in this course comes from the following response (Fig \ref{fig: table1q11}), in which $57.5\%$ of respondent voted to have hands-on approach:
\begin{figure}[H]
  \centering
  \includegraphics[width=0.45\textwidth]    {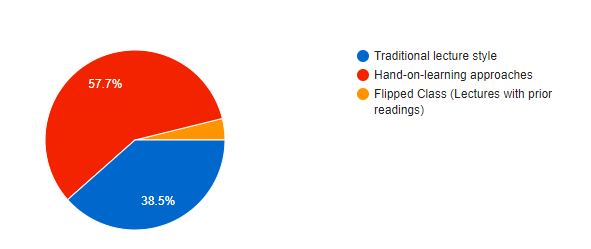}
  \caption{Response to the question: ``As a student, what teaching style do you prefer?"  from Table \ref{table: current cohort}}
  \label{fig: table1q11}
\end{figure}

Question $7$ (Fig \ref{fig: table1q7}) of Table \ref{table: current cohort} directly engages with students on their preference of Lab-based component in a RL course. Generally, more students showed their preference for labs, that is, $53.8\%$ voted for `Yes' and `Very Much' where as $46.1\%$ of students votes for `No' and `Not Very Much':
\begin{figure}[H]
  \centering
  \includegraphics[width=0.45\textwidth]    {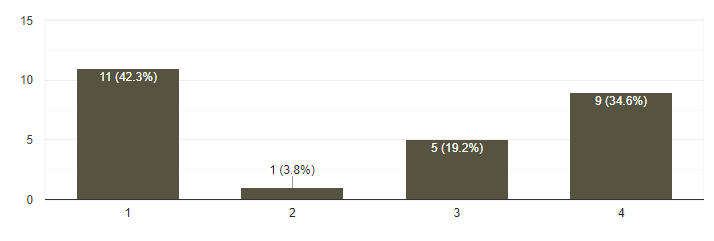}
  \caption{Response to the question: ``Would you prefer having a lab component in the RL course?"  from Table \ref{table: current cohort}}
  \label{fig: table1q7}
\end{figure}

The course evaluation feedback from the Spring 2025 semester was generally substantially more positive, with the course receiving an overall score of 4.52 out of 5. There was considerable qualitative feedback that strongly suggested positive impact of more practical elements in the course. Some of the qualitative feedback were as follows:
\begin{itemize}
    \item \textit{``Assignments were quite challenging. Classes were fun and interesting."
    \item  ``The coding assignment was a bit challenging, but it helped me learn alot"
    \item ``The coding assigments and Gymnasium were definitely the most interesting parts of the course."
    \item ``The assessments challanged different aspects of the course and increased our learning of the course."
    \item ``Learned so much about the third paradigm of Machine Learning, from theory to coding. It helped me in understanding AI better as I had already taken Deep learning."}
\end{itemize}

\subsection{Students' Performance Analysis} 
The statistical analysis of students' performance scores (expressed as percentages) from the 2023 and 2025 semesters aligns closely with the pedagogical framework proposed for undergraduate reinforcement learning (RL) education. The performance is based on the overall students' grade at the end of both semesters, which constitutes grades from quizzes, assignments and a course project. Here, descriptive statistics from Table \ref{table: Descriptive Stats} reveal that the 2025 semester, which incorporated interactive labs, achieved a higher mean score ($\mu_{2025} = 88.31$ vs.
$\mu_{2023} = 80.77$) and lower variability ($SD_{2025} = 7.92$ vs. $SD_{2023} = 16.47$) compared to the 2023 semester. This outcome supports the framework’s emphasis on integrating hands-on, lab-based learning to bridge theoretical concepts (e.g., Markov Decision Processes, Bellman equations) with practical implementation.

The Welch’s $t$-test \cite{Welch} ($t = 2.20$, $p < 0.05$) and large Cohen’s $d$ effect size \cite{cohen2013statistical} ($d = 0.59$) confirm the 2025 semester’s superior performance, validating the efficacy of structured coding exercises using platforms like OpenAI Gymnasium. These results mirror the framework’s assertion that iterative debugging and gamified learning enhance engagement and reduce variability in outcomes, as evidenced by the tighter inter-quartile range (IQR) as mentioned in Table \ref{table: Descriptive Stats}.

However, limitations persist. The analysis assumes institutional access to computational resources (e.g., TensorFlow, Gymnasium), which may exclude under-resourced settings—a concern echoed in the framework’s discussion of scalability. Additionally, the homogeneous sample (single-institution data) risks inflating effect sizes, as noted in Table \ref{table: Descriptive Stats}’s skewness values ($-1.03$ for 2025 vs. $-1.19$ for 2023), which reflect left-skewed distributions.

The non-overlapping 95\% confidence intervals ($CI_{2025}: 85.40 – 91.21 $; $CI_{2023}: 74.38 – 87.16$) in Table \ref{table: Descriptive Stats} further support the practical significance of active learning strategies. These findings resonate with the framework’s focus on scaffolding techniques (e.g., Jupyter notebooks with fading support) to foster independent problem-solving. 

While confounding variables (e.g., concurrent curricular revisions) may partially explain the 2025 cohort’s performance, the statistical evidence is significant enough to robustly supports the efficacy of proposed learning strategies. This synergy between pedagogical design and empirical validation exemplifies the potential of theory-driven innovation to advance RL education, though further refinement is needed to ensure ethical and equitable implementation across diverse educational contexts.

\section{Conclusions}
Based on the survey results and statistical outcomes robustly validate the pedagogical value of integrating theory with practice in RL education, though broader implementation requires addressing resource disparities and ethical training gaps.

\subsection{Limitations}
While the study demonstrates the efficacy of integrating interactive labs into undergraduate RL education, several limitations merit discussion to contextualize its findings and guide future research. First, the generalizability of the results is constrained by the study’s reliance on a single-institution cohort with homogeneous academic and infrastructural conditions. This limits insights into how the framework might perform in institutions with varying resource capacities, curricular priorities, or student demographics. Furthermore, the dependence on self-reported survey data introduces potential biases, such as social desirability bias, where participants may overstate positive perceptions of labs align with perceived expectations. The absence of longitudinal data also restricts the ability to assess whether observed improvements in grades and project outcomes translate to long-term skill retention or career preparedness.  Finally, while the framework emphasizes technical proficiency, ethical considerations, such as the societal implications of autonomous RL systems, are not rigorously evaluated, despite their prominence in the course’s introductory discussions. 

The reliance on tools like OpenAI Gymnasium and TensorFlow further assumes institutional access to computational resources, potentially excluding under-resourced educational settings from adopting the model. This is much more relevant to address since the students from Spring 2025 course iteration struggled in their project, where they had to train sophisticated Deep Q-Networks but on local hardware or Google Collab free version which had very limited computation resources. Platforms such as Google Collab and its higher  tier packages such as Collab Pro can help mitigate this problem given that the institution can afford the cost. The institution can also potentially collaborate with cloud-compute services such as Amazon AWS, Microsoft Azure and so on as some institutions have continued to adopt this strategy of mitigating lack of computational resources.

\subsection{Future Work}
To address these limitations, future work will prioritize scalability, methodological rigor, and ethical depth. Expanding the study to 3–5 institutions across diverse regions, including universities in the Global South and liberal arts colleges, will test the framework’s transferability and illuminate how institutional resources and cultural contexts shape outcomes. Longitudinal tracking of alumni career trajectories over 3–5 years will assess the enduring relevance of lab-acquired competencies, while comparative studies incorporating control groups will isolate the intervention’s effects from external variables. For trajectories below 3 years, relevant final year projects and students' research publications can also be a good measure of longitudinal tracking of students' long term skill retention.

Methodological enhancements will include more triangulating self-reported data with observational metrics, such as code commit frequency in labs or test-cases, to mitigate bias, alongside standardized pre/post-tests aligned with RL competencies (e.g., solving Markov Decision Processes) to quantify skill gains. These strategies will help to mitigate the issues which comes with self-reported data, thereby enhancing the rigor and credibility of data.

Ethical training will be integrated into labs through scenario-based modules, such as designing RL agents for healthcare triage or evaluating algorithmic fairness in financial systems, with outcomes assessed via reflective essays or structured debates. Ethical case scenario can be incorporated in each assessment, which could be peer-reviewed for an unbiased review, incorporating ethical nuance and ethical subjectivism.

To promote equitable access, open-source lab templates guidelines will be published on platforms like GitHub, accompanied by documentation optimized for low-resource settings (e.g., CPU-only training workflows). Concurrently, partnerships with funding bodies will ensure sustainable maintenance of these resources, fostering community-driven updates as RL techniques evolve. 

Lastly, we would encourage more researchers and educators to contribute in the development of RL education. This will bring diverse point-of-views and approaches which will ultimately lead to a more robust and cohesive pedagogical framework.

By systematically addressing these gaps, RL pedagogy will progress beyond localized pilots, establishing a globally adaptable framework that bridges theoretical rigor, practical innovation, and ethical accountability in AI education.

\subsection{Acknowledgments}
This research has been generously supported by Habib University via an Internal Research Grant.


\bibliographystyle{ieeetr}
\bibliography{main.bib}

\vspace{12pt}

\end{document}